\documentclass[%
 reprint,
 amsmath,amssymb,
 aps,
superscriptaddress]{revtex4-2}

\usepackage{graphicx}
\usepackage{booktabs}
\usepackage{dcolumn}
\usepackage{bm}
\usepackage{hyperref}
\usepackage{color}
\usepackage[normalem]{ulem}

\def\be{\begin{equation}}
\def\ee{\end{equation}}
\def\bea{\begin{eqnarray}}
\def\eea{\end{eqnarray}}
\def\pa{\partial}

\begin{document}

\preprint{APS/123-QED}

\title{How to obtain slow roll inflation driven by non-linear electrodynamics}

\author{Daniele Malafarina}%
 \email{daniele.malafarina@nu.edu.kz}
\author{Hrishikesh Chakrabarty}
\email{hrishikesh.chakrabarty@nu.edu.kz}
\affiliation{%
 Department of Physics, School of Sciences and Humanities, Nazarbayev university\\
 Astana, Kazakhstan 010000}%
\author{Ilia Musco}%
\email{ilia.musco@ung.si}
\affiliation{Center for Astrophysics and Cosmology, University of Nova Gorica, Nova Gorica, Slovenia}
\affiliation{Dipartimento di Fisica, Sapienza Università a di Roma, Piazzale Aldo Moro 5, 00185 Roma, Italy}

\date{\today}

\begin{abstract}

We establish for the first time the conditions that must be imposed on the action for a magnetic universe in a theory of non-linear electrodynamics in order to have an asymptotically de Sitter initial state followed by a slow roll inflationary phase.
We show that models so far proposed in the literature do not allow for a prolonged inflationary phase consistent with observations. We construct a Lagrangian that reduces to the Maxwell one in weak field; this is the only class of models that satisfies the required conditions for slow roll in the early Universe.

\end{abstract}

\maketitle

\section{Introduction} 

Recent observations show that the cosmic microwave background (CMB) exhibits a nearly scale invariant power spectrum \cite{Planck:2018vyg}. The most widely accepted explanation which simultaneously solves the flatness and horizon problems \cite{Guth:1980zm,Starobinsky:1980te,Linde:1981mu} requires a phase of accelerated expansion in the early Universe driven by a scalar field with a slow roll phase \cite{Planck:2018jri}. 
However scalar field inflation is not the only option, there are in fact alternative solutions which do not require a scalar field with a fine-tuned potential.
In particular it would be desirable to obtain a slow roll inflationary phase from suitable modifications of known physical fields.
Modifications of Maxwell's electrodynamics have been proposed in a variety of theoretical contexts starting from the original proposal by Born and Infeld \cite{Born:1934gh} all the way to string theory where Born-Infeld theory emerges as the lowest order term in the expansion of the bosonic open string at low-energies \cite{Fradkin:1985qd}.
Following such ideas, models for accelerated expansion driven by different theories of non-linear electrodynamics coupled to gravity have been gaining popularity \cite{Sorokin:2021tge,Altschueler:1990vp,Camara:2004ap, Novello:2003kh,Novello:2006ng,Garcia-Salcedo:2013cxa,Ovgun:2016oit,Ovgun:2017iwg,Benaoum:2022uta,Benaoum:2023ekz,Kruglov:2015fbl,Chavanis:2024qxa,Sarkar:2022jgn,Kruglov:2017vca,Kruglov:2016lqd,Kruglov:2016cdm}. 
In these models one considers a universe where the accelerated expansion phase is produced by magnetic fields of non-linear electrodynamics minimally coupled to gravity. Neglecting the bulk viscosity terms in the electric conductivity of the primordial plasma and imposing homogeneity and isotropy, one gets that only the average of the magnetic field squared $ B^2 $ is non negligible \cite{DeLorenci:2002mi}, hence the name `magnetic' Universe. 
This allows one to obtain a simple class of cosmological toy models which can be solved analytically.
Then the question is: \emph{can a theory of non-linear electrodynamics drive inflation to produce the observed CMB power spectrum?}  

In this letter we address this question by considering a general framework for a magnetic universe in General Relativity (GR) coupled to non-linear electrodynamics and determine the conditions for such a theory to produce a quasi de Sitter phase with arbitrarily small slow roll parameters.
As expected the conditions for a slow rolling inflationary phase in the early Universe are more stringent than the simple conditions for accelerated expansion. We derive these conditions here explicitly for the first time, and subsequently apply them to several proposed models of non-linear electrodynamics, showing that none of the existing proposed cosmological models satisfies these conditions. We then derive and discuss the simplest Lagrangian that does. Throughout the article we make use of geometrical units setting $G=c=1$.

\section{Evolution of a magnetic universe} 
The action for GR coupled to a theory of non-linear electrodynamics is
\begin{equation}\label{lagrangian-NLED}
    S = \int \left(\kappa R-\mathcal{L}(F)\right) \sqrt{-g}d^4x \,,
\end{equation}
where $\kappa=16\pi$, $R$ is the Ricci scalar of the Einstein-Hilbert action and $\mathcal{L}(F)$ is the Lagrangian for a theory of non-linear electrodynamics,
with $ F = F_{\mu\nu}F^{\mu\nu} $ being the electromagnetic field written in an invariant form by contraction of the Faraday tensor $ F_{\mu\nu} $. The above Lagrangian gives Maxwell's electrodynamics for $\mathcal{L}=-F/4$ with $F=2(B^2-E^2)$. Variation of the action \eqref{lagrangian-NLED} gives Einstein's equations $ G_{\alpha\beta} = 8\pi T_{\alpha\beta} $ with the energy momentum tensor due to non-linear electrodynamics given by
\begin{equation}
    T_{\alpha\beta} = -4(\partial_F \mathcal{L})F_{\rho\alpha}F^{\rho}_{\beta} + g_{\alpha\beta}\mathcal{L} \,,
\end{equation}
where $ \partial_F $ denotes a derivative with respect to $ F $.

For the cosmological model we shall consider a Friedman-Robertson-Walker (FRW) line element
\begin{equation}\label{eq-FRW}
    ds^2 = -dt^2 + a(t)^2\left[ \frac{dr^2}{1-kr^2} + r^2\left( d\theta^2 + \sin^2 \theta d\phi^2 \right) \right] \,,
\end{equation}
where $ a(t) $ is the scale factor and $k$ is the spatial curvature. 
For an arbitrary Lagrangian $\mathcal{L}(F)$ one would obtain the energy-momentum tensor as
\be 
T_{\alpha\beta}=(\rho+p)u_\alpha u_\beta +pg_{\alpha\beta}+(q_\alpha u_\beta+q_\beta u_\alpha )+\pi_{\alpha\beta},
\ee 
where $q_\alpha$ is related to the electromagnetic Poynting vector while $\pi_{\alpha\beta}$ is a symmetric, trace-free tensor which accounts for the anisotropies.

We can then obtain an isotropic FRW Universe  
when the averaged electromagnetic fields are such that
\bea \label{avg-EB1}
        \langle E \rangle &=& \langle B \rangle = 0 \,, \ \ \ \ \ \langle E_iB_i \rangle = 0 \,, \\ \label{avg-EB2}
        \langle E_iE_j \rangle &=& \frac{1}{3}E^2g_{ij} \,, \ \ \ \ \ \ \langle B_iB_j \rangle= \frac{1}{3}B^2g_{ij}, \
\eea
where the average of a quantity $X$ at a given time $t$ is defined over a sufficiently large spatial volume $V$ as
\be 
\langle X \rangle =\lim_{v\rightarrow V} \frac{1}{v}\int \sqrt{-g} X d^3x,
\ee 
with $v=\int \sqrt{-g} d^3x$.
Eqs.~\eqref{avg-EB1} imply the vanishing of the Poynting vector $q_\alpha$ while Eqs.~\eqref{avg-EB2} give isotropy of the stress-energy tensor (for more details see for example \cite{Tolman:1930ona,Barrow:1998ih,Novello:2003kh}).

Considering the form of the stress-energy tensor for a perfect fluid, i.e. $T_{\alpha\beta}=(\rho+p)u_\alpha u_\beta +pg_{\alpha\beta}$, from the Faraday tensor written in terms of the electric and magnetic fields, we then get the density and the isotropic pressures as
\bea
        \rho &=& - 4\partial_F \mathcal{L} E^2 - \mathcal{L}\,, \\
        p &=& -4\partial_F \mathcal{L} \left(\frac{2B^2 - E^2}{3}\right) + \mathcal{L} \,.
\eea
In the radiation dominated epoch of the early Universe the electromagnetic field interacts strongly with the charged plasma, effectively resulting in the screening of the electric field in the rest frame of the fluid \cite{Lemoine:1995vj}.
Therefore it is expected that the average of the electric field is negligible allowing us to set $ E^2 = 0 $ 
so that the energy density and pressures become
\bea \label{rho}
        \rho &=& -\mathcal{L}, \\ \label{p}
        p &=& \left( 1- \frac{8B^2}{3}\frac{\partial_F \mathcal{L}}{\mathcal{L}} \right) \mathcal{L} \,.
\eea
Note that the requirement for $E^2=0$ for the classical field is a consequence of the fact that we are considering scales (both in time and in space) larger than the typical scales for scattering and absorption in the primordial plasma. In this approximation the current density becomes the dominant term in the Amp\'ere equation, which allows to impose $E^2=0$ in the rest frame of the fluid.
Of course, since $\langle E_i E_j \rangle$ is a classical average, we are also neglecting quantum fluctuations which are expected to be important on shorter scales (i.e. early enough in the history of the Universe). Therefore in principle one would have to consider a full quantum treatment of the model.

The weak energy conditions, i.e. $ \rho > 0 $ and $ \rho + p > 0 $, are satisfied if we require $\mathcal{L}<0$ and $\partial_F \mathcal{L}<0$. However, looking for a non-singular cosmological model, we must allow for the energy conditions to be violated in the early Universe. 
Physically the violations of energy conditions can be understood as large scale corrections to the energy-momentum due to small scale repulsive effects \cite{Martin-Moruno:2017exc,Visser:1999de}.

We can now consider the Friedman equations, obtained from the FRW metric in \eqref{eq-FRW}, written as
\bea \label{eq-friedman}
        H^2 &=& \left(\frac{\dot{a}}{a}\right)^2 = \frac{8\pi}{3} \rho - \frac{k}{a^2}, \\
        \frac{\ddot{a}}{a} &=& -\frac{4\pi}{3} (\rho + 3p) \,,
\eea
where the dot denotes a derivative with respect to the co-moving time $t$ and $ H \equiv \dot{a}/a $ is the Hubble parameter. 
The zero component of the conservation of energy-momentum condition $ \nabla^\alpha T_{\alpha\beta} = 0 $ gives the continuity equation
\begin{equation} \label{cont}
    \dot{\rho} + 3 \frac{\dot{a}}{a}\left( \rho + p \right) = 0
\end{equation}
and it is easy to check that, for a magnetic universe (i.e. setting $E^2=0$), this equation is independent of the form of $\mathcal{L}$ and gives
\begin{equation}
    2 B +a \frac{dB}{da} = 0 \,.
\end{equation}
Once integrated this gives
\begin{equation}
    B = \frac{B_0}{a^2} \,,
\end{equation}
where $ B_0 $ is the value of the magnetic field when $a=1$ which can be taken as being today. 
One must keep in mind that $T_{\mu\nu}$ is defined in terms of the classical averages $E^2$ and $B^2$ and thus, strictly speaking, we are dealing with an averaged stress-energy tensor and therefore the Bianchi identity $ \nabla^\alpha T_{\alpha\beta} = 0 $ that gives Eq.~\eqref{cont} must be understood as $ \nabla^\alpha \langle T_{\alpha\beta} \rangle = 0 $, where $\langle T_{\alpha\beta} \rangle$ is defined in terms of the classical averages in Eqs.~\eqref{avg-EB1} and \eqref{avg-EB2}. This is a good approximation for large enough volumes although at certain scales the contributions of both classical and quantum perturbations might become important.

For simplicity we can redefine the form of the Lagrangian as
\be 
\mathcal{A}(F)=\log{\mathcal{L}(F)},
\ee 
such that $\pa_F \mathcal{L}/\mathcal{L}=\pa_F \mathcal{A} $. Then the equation of state for non-linear electrodynamics can be simply written as $p=\omega \rho$, where from Eq.~\eqref{p} we get
\be 
\omega=-1+\frac{8B_0^2}{3a^4}\pa_F \mathcal{A}=-1+\frac{4}{3}F (\pa_F \mathcal{A}) \,,
\ee 
and hence a de Sitter phase ($\omega\rightarrow -1$) is obtained when $F(\pa_F \mathcal{A})\rightarrow 0$. 
From the second Friedmann equation we can then compute the condition for accelerated expansion, i.e. $\ddot{a}>0$, as $\rho(1+3\omega)<0$ which gives
\be 
\frac{4B_0^2}{a^4}\pa_F \mathcal{A}<1,
\ee 
which is satisfied in the case of a quasi de Sitter state.

\section{Conditions for slow roll inflation} 

Let us consider a spatially flat universe (i.e. $k=0$). In order to have a quasi de Sitter period of expansion lasting long enough, the slow roll parameters defined as
\bea
    \epsilon &\equiv& - \frac{\dot{H}}{H^2}=\frac{3}{2}(1+\omega) \,, \label{epsilon1} \\
    \eta &\equiv& \frac{\dot{\epsilon}}{\epsilon H} = \frac{\omega_{,a}a}{1+\omega} \,, \label{eta1}
\eea
must be $\ll1$ as $a\rightarrow 0$, where we have set $\omega_{,a}=d\omega/da$. The condition $\omega \rightarrow -1$ is necessary in order to have $\epsilon$ small, while to achieve slow roll inflation one must impose an additional condition on $\omega_{,a}$ in order to keep $\eta$ small at the same time. 

The explicit expressions for $ \epsilon $ and $ \eta $ in any theory of non-linear electrodynamics are
\bea \label{epsilon}
\epsilon &=& \frac{4B_0^2}{a^4}\pa_F \mathcal{A} 
\,, \\
\label{eta}
\eta &=& \frac{(\pa_F \mathcal{A})_{,a}}{\pa_F \mathcal{A}}a-4 
\,.
\eea 

The requirements that $\epsilon \ll 1$ and $\eta \ll 1$ are broad enough to include bouncing models as well as asymptotically de Sitter models satisfying observational constraints. Therefore the requirement that both slow roll parameters go to zero as $a$ goes to zero asymptotically is more stringent than the simpler requirements that $\epsilon$ and $\eta$ are small. Nevertheless this is enough to exclude a large set of previously proposed models.

For example, if we assume a polynomial expansion for $\pa_F \mathcal{A}$ such that $\pa_F \mathcal{A}\sim a^n$ in the early Universe, we see that $\epsilon\sim a^{n-4}$ which goes to zero for $n>4$. However since $\eta\sim n-4$ we see that for $n>4$ the parameter $\eta$ goes to a constant value equal or larger than one.
Therefore satisfying both conditions simultaneously is not possible when $\pa_F \mathcal{A}$ is written as a polynomial expansion close to $a\simeq 0$ and the only inflationary allowed scenario in this case would be the constant roll (where $\eta$ goes to a constant) \cite{Kinney:2005vj,Motohashi:2014ppa}.

Going back to expressing $\mathcal{A}$ as a function of $F$, from Eqs.~\eqref{epsilon} and \eqref{eta} the conditions for slow roll can be expressed as conditions on $\mathcal{A}(F)$ as 
\be \label{cond-F}
F(\pa_F \mathcal{A})\rightarrow 0 \quad \text{ and } \quad  \frac{F (\pa_{FF} \mathcal{A})}{\pa_F \mathcal{A}}\rightarrow -1 \, .
\ee 

Using these within \eqref{epsilon} and \eqref{eta}, we shall analyze here a few models proposed in the literature for a magnetic universe with non-linear electrodynamics:  

{\it (1) }
The simplest example is to take Maxwell plus a term proportional to $F^n$, that is
\be 
\mathcal{L}=-\frac{F}{4}+\alpha F^n \,.
\ee
For example in \cite{Novello:2003kh,Novello:2006ng}  Novello et al. 
considered the case $n=-1$.
From the above expression we see that 
\be 
\pa_F \mathcal{A}=\frac{4\alpha n F^{n-1}-1}{4\alpha F^n-F}=\frac{a^4}{2B_0^2}\frac{4\alpha n (2B_0^2)^{n-1} - a^{4(n-1)}}{4\alpha (2B_0^2)^{n-1} -a ^{4(n-1)}},
\ee
and we cannot get a de Sitter phase for any value of $n$: 
for $n>1$ we have $\epsilon \rightarrow 2n$ while for $n<1$ we have $\epsilon \rightarrow 2$.

{\it (2) } A second model worth investigating is the Born-Infeld Lagrangian \cite{Born:1934gh}
\be 
\mathcal{L}=\alpha\left(1-\sqrt{1+\frac{F}{2\alpha}}\right) \, ,
\ee 
which gives 
\be 
\partial_F\mathcal{A}= -\left[4\alpha\sqrt{1+\frac{F}{2\alpha}}\left(1-\sqrt{1+\frac{F}{2\alpha}}\right)\right]^{-1}\, ,
\ee 
for which we get $\epsilon\rightarrow 1/2$. We conclude that a magnetic universe in the Born-Infeld theory does not have an initial de Sitter phase.

{\it (3) } In \cite{Kruglov:2015fbl,Chavanis:2024qxa} the following Lagrangian was proposed
\be 
\mathcal{L}=-\frac{F}{1+F/\alpha}
\ee 
and we see that for a magnetic universe with $F=2B^2$ we have
\be 
\pa_F \mathcal{A}=\frac{\alpha}{F(\alpha+F)}= \frac{\alpha a^8}{2B_0^2(\alpha a^4 + 2B_0^2)} \,.
\ee 
While the condition for a de Sitter initial phase is satisfied, for the slow roll parameters we get
\be
\epsilon = \frac{2 a^4}{a^4+2B_0^2/\alpha}, \ \ \ \  \eta = 4-2\epsilon,
\ee
and thus even though $\epsilon \rightarrow 0$ when $a\to0$ we have that $\eta \rightarrow 4$, implying constant roll.

{\it (4) } Another alternative Lagrangian was considered in \cite{Ovgun:2017iwg}:
\be 
\mathcal{L}=-\frac{Fe^{-\alpha F}}{\alpha F+\beta} \,,
\ee 
for which we get
\be 
\pa_F \mathcal{A}=\frac{\beta a^8}{2B_0^2(\beta a^4+2\alpha B_0^2)}-\alpha \,,
\ee
which does not give a de Sitter phase. Similar expression for the Lagrangian were considered in \cite{Sarkar:2022jgn,Kruglov:2017vca}, although it should be noted that the calculations for $\epsilon$ and $\eta$ in \cite{Sarkar:2022jgn} are not correct; while in \cite{Kruglov:2017vca} the derivation of the spectral index and the tensor to scalar ratio is not reliable, because it does not implement the full perturbation analysis. Another Lagrangian with $\mathcal{L}=-Fe^{-\alpha F}$ was studied in \cite{Kruglov:2016lqd} and a model with $\mathcal{L}=-\alpha \arctan (F/\alpha)$ was proposed in \cite{Kruglov:2016cdm}. It is easy to check that the conditions in Eq.~\eqref{cond-F} are not met by any of these models.

{\it (5)} Finally a commonly used Lagrangian for achieving regular black holes was presented in \cite{Fan:2016hvf,Toshmatov:2018cks,Bronnikov:2017tnz,Malafarina:2022oka}:
\be
    \mathcal{L} = \frac{\lambda}{4\alpha}\frac{\left( \alpha F \right)^{(k+3)/4}}{\left[ 1 + \left( \alpha F \right)^{k/4} \right]^{(k+\lambda)/k}},
\ee 
where $\alpha$ is a dimensional coupling constant and $k\geq 1$ and $\lambda \neq 0$ are the model's parameters. Maxwell's electrodynamics is recovered for $k=1$ and $\lambda=-1$, while the black hole singularity is resolved by taking $\lambda\geq 3$ and the Hayward regular black hole is obtained by setting $k=\lambda=3$ \cite{Malafarina2023}.
From
\be 
\pa_F \mathcal{A}= \alpha a^4\frac{(k+3)a^k+(3-\lambda)\beta_0^{k/4}}{4\beta_0(a^k+\beta_0^{k/4})},
\ee 
we get
\bea 
\epsilon &=& \frac{(k+3)a^k+(3-\lambda)\beta_0^{k/4}}{2(a^k+\beta_0^{k/4})},   \\
\eta &=& ka^k\frac{(k+\lambda)\beta_0^{k/4}}{(a^k+\beta_0^{k/4})\left[(k+3)a^k+(3-\lambda)\beta_0^{k/4}\right]},
\eea 
where $\beta_0 = 2\alpha B_0^2$. It is easy to see that for $a\rightarrow 0$ we have $\epsilon\rightarrow (3-\lambda)/2$. Therefore to satisfy the first condition we must fix $\lambda=3$. However $\eta \rightarrow 0$ for every value of $k\geq 1$ when $\lambda \neq 3$, while for $\lambda =3$ we have that $\eta \rightarrow k \neq 0$. Therefore it is not possible to satisfy both conditions for any model belonging to this class.

It is important to notice that the coupling constant $\alpha$ (as well as other constants in front of $F$) does not play any role in the behavior of the slow roll parameters. The only free parameters that can affect the limiting values of $\epsilon$ and $\eta$ as $a$ goes to zero appear in the powers of $F$ (or functions of $F$). Therefore, unless one is willing to consider for example terms like $F^\gamma$ with $\gamma\simeq 10^{-3}$, the conditions in Eq.~\eqref{cond-F} effectively rule out these models as candidates for inflation. Allowing such term is what was done for example in \cite{Benaoum:2022uta} where the authors considered the Lagrangian $\mathcal{L}=-F/(\alpha F^\gamma+1)^{1/\gamma}$ and must fix $\gamma\simeq 0.01$ in order to obtain $\eta=4\gamma \ll 1$.

\section{Magnetic Universe with slow roll inflation}

\begin{figure*}
    \begin{center}
        \includegraphics[width=8.5cm]{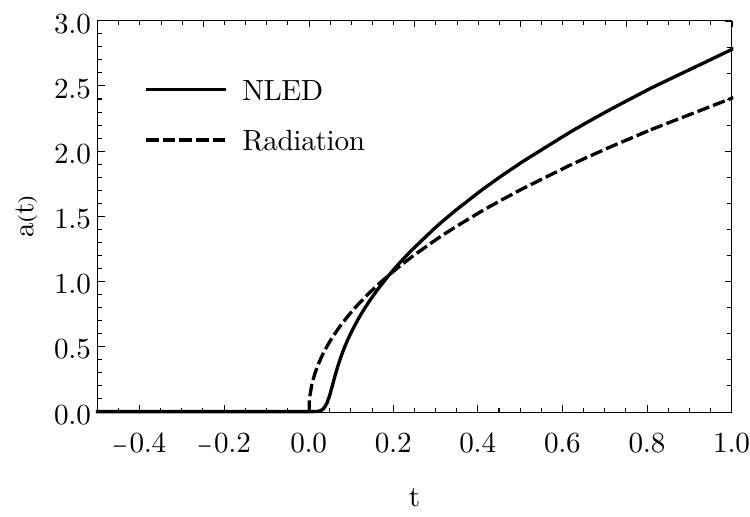} \hspace{0.5cm}
        \includegraphics[width=8.5cm]{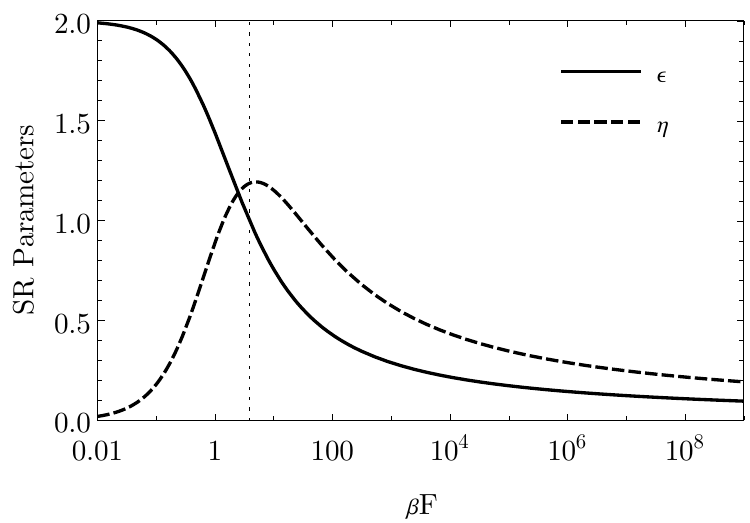}
    \end{center}
    \caption{ Left: Comparison of the time evolution of the scale factor $a$ between a magnetic universe dominated by the nonlinear
electrodynamics Lagrangian as in Eq.~\eqref{our} (solid line) and a radiation dominated universe (dashed line). The values of $\beta = 10^{-3}$ and the strength of the magnetic field today $B_0 = 2$, are chosen for illustrative purposes. Right: Evolution of the slow-roll parameters with respect to $\beta F$ for the Lagrangian in Eq.~\eqref{our} (notice that $\beta F$ grows as $t$ goes to minus infinity). In the early universe, i.e. for $\beta F$ large, both parameters go to zero. the vertical dotted line represents the end of inflation.  \label{fig-a}}
\end{figure*}

From the above considerations we can summarize the necessary conditions for the action of non-linear electrodynamics to construct a Lagrangian producing slow roll inflation.    
Let us define a function $f(a)$ as
\be 
\pa_F \mathcal{A}=a^4 f(a) ,
\ee 
so that finding a theory with arbitrarily small slow roll parameters, reduces to the simpler problem of finding a function $f(a)$ for which 
\be \label{cond}
\lim_{a\rightarrow 0} f(a)=0 \quad \text{ and } \quad \lim_{a\rightarrow 0} \frac{af'(a)}{f(a)}=0 \,.
\ee

A suitable function that satisfies both conditions in \eqref{cond} is given by
$f(a) \simeq (\log a^4)^{-1}$,
for which the Lagrangian becomes
$\mathcal{L} = -\alpha\log F$,
where $\alpha$ is some constant. For such a theory of non-linear electrodynamics it is possible to achieve a quasi de Sitter initial period as $a\rightarrow0$ and we can easily show that both the slow roll parameters $\epsilon$ and $\eta$ are small in the same limit. However, this Lagrangian does not have a Maxwell limit. 

 The most general Lagrangian satisfying the conditions on $\epsilon$ and $\eta$ and  possessing a well-behaved Maxwell limit for large $a$ can be written as
\be\label{our}
\mathcal{L} = - \alpha \log \left( 1 + \beta F + h(F) \right),
\ee
with $h(F)$ being an arbitrary function that goes to zero for $F\ll1$ and $F\gg1$, and where we must set $\alpha>0$ and $\beta>0$ to respect causality \footnote{This can easily be seen from the energy conditions with the requirement $\alpha>0$ necessary to ensure positive energy density and the requirement $\beta>0$ imposed to fulfill $\rho+p\geq 0$.} and $\alpha(\beta+ h'(0))=1/4$ to recover Maxwell's theory in the weak field. Notice that the argument of the $\log$ must go linearly in $F$ in the weak field to recover Maxwell theory and in the strong field to satisfy the slow roll conditions. Then the function $h(F)$ affects the evolution of the Universe only between the end of inflation and late times. It can also be proved that this is the only class of Lagrangians with a Maxwell weak field limit that satisfies both conditions. 

The easiest example to consider is taking $h(F)=0$: this theory was originally introduced in \cite{Soleng:1995kn} in the context of black holes and studied in more detail in \cite{Gaete:2018nwq, Kruglov:2014iqa} but was not applied to cosmology. Interestingly the same Lagrangian was also obtained from an entirely different starting point in \cite{Zholdasbek:2024pxi}. It is also interesting to note that the singularity in this model is shifted to past infinity. In fact the Kretschmann scalar $\mathcal{K}$ goes as $ \rho^2$ for small $a$ and diverges as $a$ goes to zero. 
In Fig.~\ref{fig-a}, we illustrate the qualitative behavior of the cosmological model with the Lagrangian in Eq.~\eqref{our}, with $h(F)=0$. The left panel shows the scale factor plotted against time, comparing the evolution of a magnetic universe, where $a\to0$ for $t\to-\infty$, to a spatially flat universe dominated by radiation, where the big-bang singularity appears at $t=0$. 

Now this Lagrangian is well-behaved in the low energy limit as it reproduces standard electrodynamics and it also satisfies both of the conditions in Eq.~\eqref{cond} since
\bea \label{epsilon-model}
\epsilon &=& \frac{2\beta F}{(1+\beta F)\log (1+\beta F)} \,,   \\ \label{eta-model}
\eta &=& 4 \frac{\beta F-\log(1+\beta F)}{(1+\beta F)\log(1+\beta F)} \,.
\eea  
Notice that in the Maxwell limit $F\simeq 0$ we retrieve the expected values of both parameters. On  the other hand for $F$ large, i.e. when $a$ goes to zero, we now have that both slow roll parameters are arbitrarily small. 
The Lagrangian in Eq.~\eqref{our} is the simplest model of non-linear electrodynamics which satisfies both conditions and is therefore a suitable candidate for describing inflation, as an alternative to scalar fields. 
In the right panel of Figure \ref{fig-a} we show the behavior of the slow roll parameters plotted against $\beta F$, where time increases for decreasing values of  $\beta F$. 

The asymptotically de Sitter phase allows one to easily fulfill the requirement that inflation lasts for the required number of e-foldings, as one just needs to set the initial time of inflation accordingly. Then, provided we know the strength of the magnetic field at the end of inflation, the scale of inflation can be calculated from the required number of e-foldings. 
The end of inflation is achieved when $\ddot{a}=0$, or equivalently when $\epsilon=1$, and from Eq.\eqref{epsilon-model} we see that $\epsilon=1$ approximately when $\beta F \simeq 3.9$, while the Maxwell limit $\epsilon \rightarrow 2$ is obtained for $\beta F \rightarrow 0$. The end of inflation is shown with a vertical dotted line, and in this particular model ($h(F)=0$) is very close to the maximum value of $\eta$. 
This shows that inflation terminates before the fields becomes Maxwell-like, suggesting that the only free parameter $\beta$ (remember that $\alpha$ must be set in order to obtain the Maxwell limit in the weak field) may be chosen in such a way to fulfill requirements for a magnetically dominated reheating to be compatible with existing constraints on primordial magnetic fields \cite{Saga:2017wwr}.

Testing the validity of this model against observations is beyond the scope of the present letter. In the future we plan to evaluate the perturbations to calculate the power spectrum of tensor and scalar fluctuations which can be constrained by observations of the CMB.

\section{Discussion} 

Theories of non-linear electrodynamics have been used extensively to construct regular black hole solutions 
\cite{Ayon-Beato:1998hmi, Bronnikov:2000vy, Dymnikova:2004zc, Fan:2016hvf, Malafarina:2022oka} (it is important to notice that, similarly to other regular black hole proposals, such solutions are often unstable \cite{DeFelice:2024seu}). It is then natural to ask whether the same strong field effects may have played a role in the early Universe, providing an alternative to the big-bang singularity, and if they may be used in place of scalar field inflation to obtain the observed properties of the CMB power spectrum. We have shown that while it is relatively easy to achieve a quasi de Sitter phase of accelerated expansion in a non-linear magnetic universe, the same can not be said for an inflationary phase with slow roll parameters that produce a scale invariant CMB power spectrum. We have reduced the conditions on the slow roll parameters to two simple conditions on the Lagrangian for non-linear electrodynamics and we have shown how the examples previously proposed in the literature do not satisfy both conditions. Of course constant roll inflation models may be useful to explain other features of the early Universe such as the production of primordial black holes \cite{Ivanov:1994pa,Inoue:2001zt,Motohashi:2017kbs,Motohashi:2019rhu}. 
Also it may be possible to achieve slow roll inflation by including additional elements such as non vanishing electric fields, the dual of the Faraday tensor or non minimal coupling to gravity (see for example \cite{Otalora:2018bso,Avetisyan:2021heg,Campanelli:2007cg}).
From the proposed conditions we constructed a theory that was never discussed before in this context and produces the desired effects in the early Universe. 
We note that a realistic assessment of the model’s viability and detailed investigation of the perturbations and evaluation of the spectral index and tensor to scalar ratio in the proposed theory would require addressing the quantum nature of the underlying vector fields. In particular, the renormalization of the expectation value of $\langle T_{\mu\nu} \rangle$ in a de Sitter background derived from nonlinear
electrodynamics poses nontrivial challenges. A complete quantum treatment of these aspects is beyond the scope of the present classical analysis and will be the subject of future work.

\section*{Acknowledgement} 

DM would like to thank INFN and Sapienza University in Rome for the warm hospitality during the preparation of the manuscript. 
The authors wish to thank Antonio Iovino and John Miller for useful comments and discussion.
DM and HC acknowledge support from Nazarbayev University Faculty Development Competitive Research Grant Program No. 040225FD4737.
The work of IM has been partially supported by the MUR PRIN Grant 2020KR4KN2 ``String Theory as a bridge between Gauge Theories and Quantum Gravity'', by the FARE programme (GW-NEXT, CUP:~B84I20000100001), and by the INFN TEONGRAV initiative. 
This publication is has received funding from the European Union’s Horizon Europe research and innovation program under the Marie Sklodowska-Curie COFUND Postdoctoral Programme grant agreement No. 101081355-SMASH and from the Republic of Slovenia and the
European Union from the European Regional Development Fund. 

Co–funded by the European Union. Views and opinions expressed are however those of the author(s) only and do
not necessarily reflect those of the European Union or European Research Exacutive Agency. Neither the European Union nor the granting authority can be held responsible for them.


\end{document}